\documentclass{iaus}
\usepackage{graphicx}

\title[The dwarf LSB population in different environments] 
{The dwarf low surface brightness population in different environments of the Local Universe}

\author[Sabatini S. et al]   
{S. Sabatini$^1$%
  \thanks{Present address: INAF-OAR, via Frascati 33, Monte Porzio Catone, Roma,Italy},
 J.I. Davies $^2$, S. Roberts$^2$, R. Scaramella$^1$}

\affiliation{$^1$ INAF-Osservatorio Astronomico di Roma, via Frascati 33, 00040 Monte Porzio Catone, Roma,Italy
\break email: sabatini@oa-roma.inaf.it; kosmobob@oa-roma.inaf.it \\[\affilskip]
$^2$School of Physics and Astronomy, Cardiff University, 5 The
Parade, Cardiff, CF24 3AA, UK \break email:
Jonathan.Davies@astro.cf.ac.uk;
Sarah.Roberts@faulkes-telescope.com}

\pubyear{2007}
\volume{244}  
\pagerange{1--10}
\date{?? and in revised form ??}
\jname{Dark Galaxies and Lost Baryons IAU Symposium}
\editors{A.C. Editor, B.D. Editor \& C.E. Editor, eds.}
\begin{document}

\maketitle

\begin{abstract}
The nature of the dwarf galaxy population as a function of
location in the cluster and within different environments is
investigated. We have previously described the results of a search
for low surface brightness objects in data drawn from an East-West
strip of the Virgo cluster (\cite{sabatini03}) and have compared
this to a large area strip outside of the cluster
(\cite{roberts04}). In this talk I compare the East-West data
(sampling sub-cluster A and outward) to new data along a
North-South cluster strip that samples a different region (part of
sub-cluster A, and the N,M clouds) and with data obtained for the
Ursa Major cluster and fields around the spiral galaxy M101. The
sample of dwarf galaxies in different environments is obtained
from uniform datasets that reach central surface brightness values
of ~26 B mag/arcsec$^2$ and an apparent B magnitude of 21
(M$_B$=-10 for a Virgo Cluster distance of 16 Mpc). We discuss and
interpret our results on the properties and distribution of dwarf
low surface brightness galaxies in the context of variuos physical
processes that are thought to act on galaxies as they form and
evolve. 
\end{abstract}

\firstsection 
\section{Introduction}
Although dwarf galaxies are the most abundant galaxies of the
Local Universe, their number counts in the field (measured by the
faint end part of the Luminosity Function, LF, \cite{blanton01},
\cite{norberg02}) are well below those predicted by the Mass
Function (MF) of $\Lambda$CDM hierarchical galaxy formation models
(\cite{kauffman93},\cite{klypin99},\cite{moore99}) - this is known
as the 'missing satellites' problem. However, recent discoveries
of ultra-faint dwarfs of the Local Group made with SDSS data
(\cite{irwin07} and references therein) have doubled the number of
known dwarf satellites of the MW, reducing the discrepancy between
observations and predictions, even if a factor of 4 is still
missing in order to solve the missing satellites problem
(\cite{simon_geha07}). These results prove that, even in the Local
Universe, the census of the dwarf galaxy population is still
incomplete and the slope ($\alpha$) of the faint end part of the
LF is therefore still subject to revision: selection effects due
to the intrinsic faint total magnitude and low surface brightness
of dwarf galaxies, have made them particularly elusive in surveys
that were not purposely designed to select them. On the
observational side, the issue is further clouded by non-uniform
datasets: different surveys reach different magnitude and surface
brightness limits and detection methods and selection criteria
vary in the way they identify dwarf galaxies. This can lead to
studies of the LF in which different authors report extremely
different results even for the same region of sky: see for example
the Fornax cluster where an extremely steep LF ($\alpha \sim
-2.0$) is found by \cite{kambas00}, while a flat one ($\alpha \sim
-1.1$) is reported by Mieske et al (this conf). In this talk we
report on our ongoing work to try and quantify the numbers of Low
Surface Brightness (LSB) dwarf galaxies as a function of the
environment in the Local Universe (within ~20 Mpc) and to assess
and compare their properties. We reduced as much as possible the
observational biases with the uniformity of our data and by using
exactly the same detection algorithm and selection criteria for
identifying dwarf galaxies in all the different environments that
we analysed. To reach this goal we put very strong requirements on
the way we carried out our work:
\begin{itemize}
\item the dataset and the data analysis need to be homogeneuos. As
already mentioned, data in the literature are highly
inhomogeneous: often the detection methods, candidate selection
and the resulting completeness of the compared samples are
different. In order to avoid these problems, our data are obtained
in exaclty the same way for the different environments (same
observing instrumentation and setup, same data analysis, detection
and selection method); \item data need to go deep both in
magnitude and surface brightness. Flux limited surveys sample only
our neighborhoods for the dwarfs contribution and easily miss
further away denser environments like the Virgo Cluster; surface
brightness limited surveys miss the contribution by LSB galaxies
(e.g. the SDSS catalogue misses 50\% of galaxies from
$\mu_{50,r}\simeq 23.5$ mag/arcsec$^2$, \cite{blanton05}). our
method allows the detection of galaxies with intrinsic properties
of 23$\leq$$\mu_{o}$$\leq$26 B mag/arcsec$^2$ and
-10$\geq$$M_{B}$$\geq$-14, where $\mu_{o}$ and $M_{B}$ are the B
band exponential central surface brightness and absolute magnitude
respectively ; \item data need to cover different environments:
inhabiting the shallowest potential wells and thus being the most
sensitive to internal and external physical processes that control
galaxy formation, dwarfs are the best test-beds to investigate how
the environment affects galaxy evolution and to estimate the
contribution of different physiscal processes (photo-ionization,
tidal interactions, SN winds, ram pressure stripping, harassment).
We obtained data from different environments of the Local
Unieverse (the field, M101, the Ursa Major Cluster, the Virgo
Cluster). Given the magnitude limits of our survey (see previous
point), the galaxies that we analyse are fainter than those
typically detected in the comprehensive survey of the Virgo
cluster carried out by \cite{binggeli84} and extends down towards
the properties of the 'classical' dSph galaxies of the Local Group
(excluding the recently discovered ultra-faint
dwarfs). 
\end{itemize}

\section{Data and data analysis}\label{sec:data}
The optical data were obtained using the Wide Field Camera (WFC)
on the Isaac Newton Telescope, La Palma, Canary Islands as part of
the Wide Field Survey (WFS), a multicolour data survey covering
over 200 deg$^{2}$ of sky. The WFC is a mosaic of four thinned EEV
4K $\times$2K CCDs with a pixel size of 0.33 arc sec and total sky
coverage of 0.29 deg$^{2}$. Images on CCD 3 were not used due to
its vignetting, this reduced our total field of view to 0.21
deg$^{2}$. The Virgo data, taken during observing runs in 1999 and
2002, consists of two perpendicular strips extending from the
centre of the Virgo cluster (defined as M87) outwards by 7 deg and
covering a total area of about 28 $\deg^2$.
\begin{figure}
\centerline{
\includegraphics[height=2in]{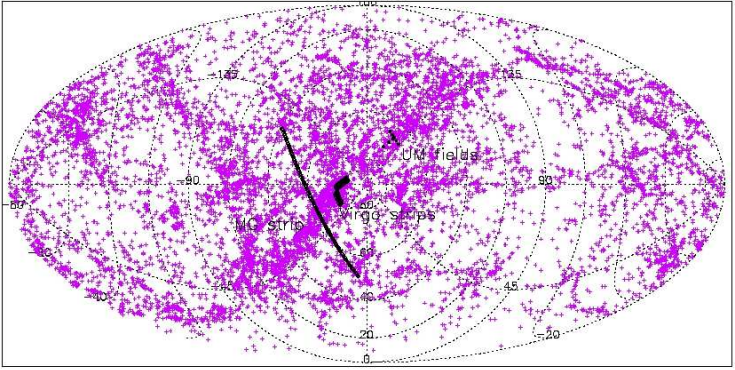}
}
  \caption{
  Position of Virgo Cluster data strips, Ursa Major and MGS data viewd from the
  North galactic pole. Also plotted are all galaxies listed in NED with v< 4500 km/s.
}\label{fig:datamap}
\end{figure}
These 2 data strips sample different regions of the Virgo cluster,
with one lying roughly perpendicular to the super-galactic plane
(E-W) and covering sub-cluster A and outward, and the other (N-S)
almost parallel to it, sampling clouds N,M. The Ursa Major data
was obtained in 2002: they consist of 8 fields going along a N-S,
E-W cross in the cluster and totalling 1.68 $\deg^2$. The M101
data was obtained in 2004 and consist of 72 frames covering an
area of just over 15 $\deg^2$, in a box around M101. The field
region is taken from the Millennium Galaxy Strip (MGS) and
consists of an area of 30 $\deg^2$ along the celestial equator
(\cite{liske03}), passing through filaments and voids and crossing
the Virgo Southern extension at its mid point. The different
environments of this survey are shown in black in Fig
\ref{fig:datamap}. All B band exposures were for 750s, whilst the
exposure time for the i band images was 1000s. All data reduction
was carried out by the Cambridge Astronomical Survey Unit pipeline
(http://www.ast.cam.ac.uk/~wfcsur/index.php). This included
de-biasing, bad pixel replacement, non-linearity correction,
flat-fielding, de-fringing, gain correction and photometric
calibration.

\subsection{Data analysis}
The detection of LSB dwarf galaxies is subject to strong selection
effects, due to their intrinsic low surface brightness that is, by
definition, below the sky level (i.e. they consist of low
signal-to-noise objects on a digital image). For this reason,
standard algorithm based on a connected pixels above a threshold
concept (e.g. SExtractor, \cite{bertin96}) are not ideal for their
detection. The galaxy detection in this work was therefore
performed using a purposely built, fully automated, detection
algorithm based on the convolution of the original images in the
Fourier Space with a set of multi-scale filters of exponential
profile. The method and its potentialities are discussed in
Scaramella et al (this conf). The detection algorithm produces a
list of candidate LSB galaxies that need to be selected against
the background galaxies. In order to estimate the best selection
criteria (i.e. the one that optimize cluster members detection and
minimize background contamination of the sample), we ran numerical
simulations of a cone of Universe uniformly populated by galaxies
and of a cluster at the position of the Virgo Cluster. The details
of this detection algorithm and the selection criteria have been
extensively discussed in Sabatini et al. (2003) and Roberts et al.
(2004). The result of this process is that we selected galaxies
that, at the distance of the Virgo cluster (16 Mpc,
\cite{jerjen04}), satisfy the following criteria: a central
surface brightness of $23 \leq \mu_{o} \leq 26$ B mag/arcsec$^2$,
and an exponential scale size ($h$) in the range: $4 \leq h \leq
9$ arcsec. \footnote{This actually corresponds to galaxies with
intrinsic scale sizes of 3 arcsec as it can be shown that
convolution of the typical seeing function of our data with a 3
arcsec scale length resulted in a measured scale length of about 4
arcsec (\cite{sabatini03}).} Our final sample consist of 596 dwarf
LSB galaxies (roughly 2/3 out of which were previously
uncatalogued) found over an area of ~70 deg$^2$ in environments of
increasing density.

\section{Results}\label{sec:results}
\subsection{Number density and Dwarf-to-Giant Ratio}
In the EW strip of the Virgo Cluster, we found on average about 18
galaxies per deg$^2$ but ranging from about 40 per deg$^2$ at the
cluster centre to approximateley 4 per deg$^2$ at the cluster edge
(\cite{sabatini03}). This decline in numbers with cluster radius
is a good indicator that we have selected a predominately cluster
population rather than background galaxies. In a related paper
(\cite{roberts04}) we compared the Virgo cluster result with field
data from the MGS. Over this region the average number of galaxies
detected were only 4 per deg$^2$, consistent with the numbers we
detected at the edge of the Virgo cluster, again indicating that
our cluster data consists predominately of cluster galaxies.
Comparing the numbers of dwarf galaxies to that of the giant
galaxy population over the same region of sky gave a
Dwarf-to-Giant Ratio (DGR) of about 14 (DGR=Num($M_{B}\leq
-19$)/Num($-14 \leq M_{B} \leq - 10$)) for the Virgo E-W strip and
22 for the N-S strip. The DGR of the MGS strip was at most 6,
consistent with the numbers of dwarf and giant galaxies found in
the Local Group (Mateo, 1998) if we imagine to move the Local
Group to the Virgo Cluster distance and apply the above discussed
selection criteria. The Virgo cluster clearly supports many more
dwarf galaxies per giant than do galaxies in less rich
environments. Values of DGR are shown in table \ref{tab:prop}
(column 3) for the environments where it was possible to calculate
them. Average number density of dwarf galaxies is also shown
(column 2) and can be compared in different environments. The MGS
survey clearly shows that there is no large field population of
LSB low luminosity galaxies that would steepen the luminosity
functions of \cite{blanton01} and \cite{norberg02} at fainter
magnitudes, to account for the discrepancy with theoretical
predictions. In the special environment of the Virgo cluster,
consistently with previous shallower surveys, we found relatively
large numbers of dwarf galaxies, that provide a little closer
match to the theory (Moore et al., 1999). \\
\subsection{Properties of the dwarf galaxies in our sample}

\begin{figure}
\begin{centering}
\includegraphics[height=2in]{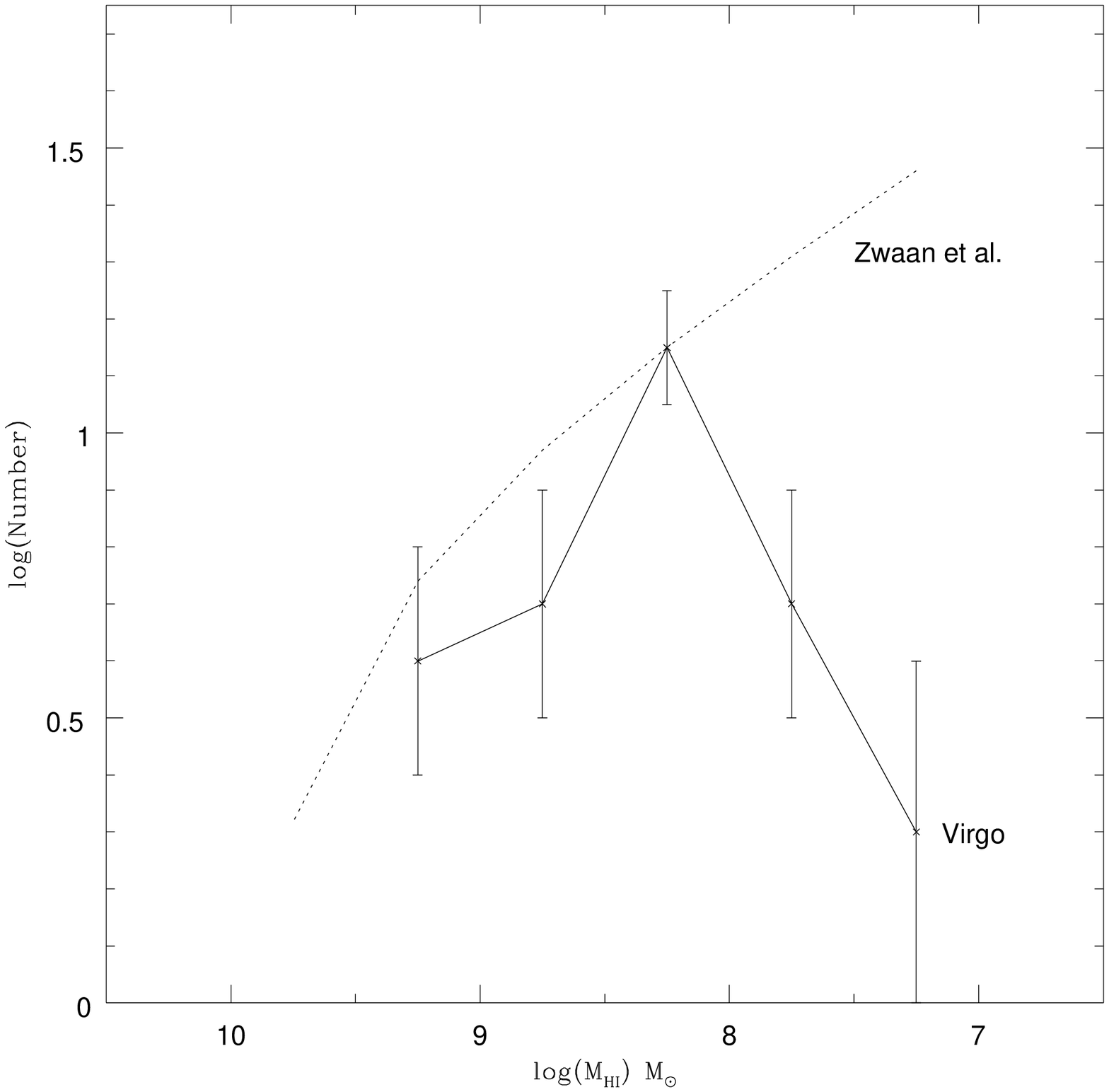} \includegraphics[height=2in]{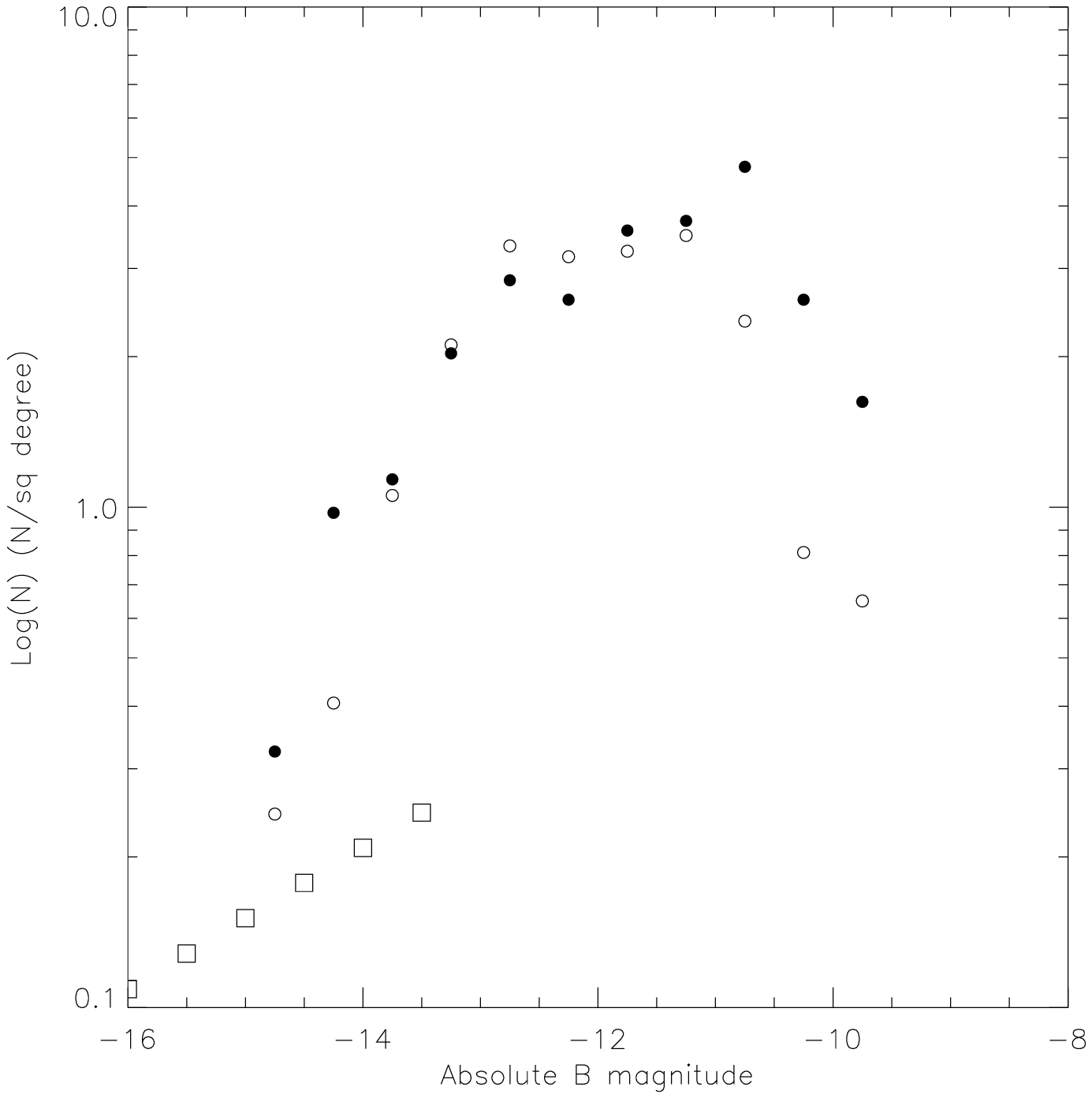}
  \caption{
  On the left: HI mass function. The
  solid line shows the mass function from the Virgo HIPASS data,
  the dashed line is the derived fit to the field galaxy mass
  function from \cite{zwaan05}, arbitrarily normalised to the peak
  on the Virgo data at $M_{HI}=8.5$. On the right: Virgo faint end
  part of the LF. Open circles are counts for the raw data and
  filled points are those corrected for detection efficiency and
  background contamination. Squares are points from the VCC LF
  with a slope of -1.3.
}\label{fig:himflf}
\end{centering}
\end{figure}

In \cite{sabatini05} we followed up the galaxies of our catalogue
and analysed the B-I colours and gas content (through observations
of our candidates carried out at Arecibo, with a mass limit of
$\sim 1.5 \times 10^7 M_{\odot}$ at the distance of the Virgo
Cluster and a column density limit of $\sim 5 \times 10^{18}
atoms/cm^2$). Table \ref{tab:prop} shows average values of the B-I
colours (column 4). Interestingly colours of the two different
regions analysed in Virgo are similar, but the NS strip shows a
higher dispersion: this could be due to the strip being populated
by galaxies from further away infalling clouds (implying fainter
magnitudes and therefore larger errors in the
colour)\footnote{This picture is also supported by the comparison
of the total magnitude distributions of the two
sample and the surface brightness (SB) distributions: 
galaxies in the NS strip have fainter total apparent magnitude and
than galaxies in the EW strip similar SB, suggesting that they are
at a higher distance (\cite{roberts07})}. On average dwarfs in
Virgo are bluer than typical giant ellipticals in the cluster
((B-I)$= 2.1\pm 0.1$, \cite{michard00}) and wihin the errors have
similar (B-I) colour to typical globular clusters (GC) around M87
((B-I)$=1.65 \pm 0.25$, \cite{conture90}). It's also worth
noticing that (B-I) colours of galaxies in the 2 cluster strips
does not depend upon distance from
the cluster centre. \\
HI detection rate for a subsample of galaxies in our surveys of
the Virgo Cluster and the field that were followed up are shown in
table \ref{tab:prop} (column 5): galaxies in the cluster tend to
be gas poor compared with those in the field. In aggrement with
this picture, although still under intense discussion at the low
masses (see this conf), we find that the HI Mass Function (HIMF)
in the cluster is shallower than in the field (see Fig
\ref{fig:himflf}). This has an opposite trend to the LF, that we
find steep in Virgo and flat in the field (Fig \ref{fig:himflf}),
suggesting that the cluster plays an important role in promoting
star formation and therefore in converting atomic gas into stars. \\
To summarise: in the comparison of Virgo cluster dwarf galaxies to
MGS field dwarf galaxies, consistent with previous observations,
we find that the cluster population is redder and gas poor.  \\
Concerning the morphology of galaxies in the different
environments analysed, again consistent with previous results in
the literature, we find that dE are more abundant in denser
environments and tend to be a small percentage in the field (table
\ref{tab:prop}, columns 6 to 8).

\begin{table}\def~{\hphantom{0}}
  \begin{center}
  \caption{Comparison of dwarf galaxy's properties in different environments}
  \label{tab:prop}
  \begin{tabular}{llllllcccccccc}\hline
      $Region$  & $<Density>$   &   $DGR$ & $<B-I>$ & $HI$ & $\% dE$ & $\% dIrr$ & $\% Other$ \\
       $$ & (N/deg$^2$) & $$ & $$ & (det. rate) & $$ \\ \hline
       MGS & 4 &   $<6$& 1.2$\pm$0.3 &$23\%$ &  24 \%   &  33 \% &   43 \%\\
       M101* &  3.6 & $-$&  $-$ & $-$&  40 \%   & 38 \%  &  22 \% \\
       M101  &  0.1 & $1$ & $-$  &$-$ & 100 \%  & $-$  & $-$  \\
       UMa & 4 & $-$&  $-$  & $-$ & 33 \%  &  17 \%   & 50 \% \\
       Virgo EW  &  18 & $14$& 1.8$\pm$0.4  & $5\%$& 62 \%   & 26 \% &   12 \% \\
       Virgo NS  &  22 & $22$& 1.8$\pm$0.7 & $-$& 54 \%  &  22 \%  &  24\% \\
       \hline
  \end{tabular}
 \end{center}
\end{table}

\section{Discussion}
In what follows we will discuss the properties of these extremely
faint low surface brightness galaxies in relation to the
environment in which they reside and evaluate which of the
analysed physical mechanisms can be the main driver for their
evolution. The main issue that we are trying to assess is wheter
these galaxies are fossils of the primordial fluctuations
predicted by $\Lambda$CDM models or they have been created more
recently.
\cite{kauffman93} suggest that the missing DM halos must remain
dark and so undetectable. If this is so, then a much larger
fraction of field DM halos remain dark compared to those in
clusters. The cluster environment must have played an important
role in making these galaxies visible. Alternately if these dark
halos do not survive to the present epoch, the cluster environment
must have created the excess of dwarf galaxies compared to the
field. In what follows we analyse mechanisms in favor of one or
the other scenario.
\subsection{Creation of dwarfs}
\begin{itemize}
\item {\bf Harassment -} \cite{moore99} showed that infalling LSB
disk galaxies in a cluster can be morphologically transformed into
dEs by the several high speed tidal interactions with giant
cluster galaxies. We tested 2 predictions of this model:\\
1) \emph{size} of the dEs in our sample. Moore et al. (1998) give
the smallest radius of an harassed galaxy to be $\sim$1.67 Kpc,
even if \cite{sabatini05} calculated that if the dwarf galaxies
found in the cluster originally came from a population of larger
field galaxies then they should have tidal radii of the order
$\sim$ 7kpc ($r_t\simeq R_{core} \sigma_{gal}/\sigma_{clus}$). The
dwarf galaxies in our sample have scale-lengths between 4 arcsec
and 9 arcsec, which, at the distance of Virgo correspond to
physical scale-sizes of 0.25kpc to 0.75kpc, therefore smaller than
the predicted sizes.\\
2) \emph{tidal streams} (i.e. stars torn out from the harassed
galaxies will lie along narrow streams which follow the orbital
path of the galaxy). \cite{davies05} carried out a search for
these tidal arcs around 38 dE galaxies found in our survey of the
N-S strip, reaching the limit of $\sim$29 B mag/arcsec$^2$ (which
is 1.5 mag fainter than the predicted surface brightness values
for the stream stripped by an infalling $M_B=20$ galaxy
distributed around a radius of 1 Mpc and 2 Kpc thick,
\cite{moore96}). From this search, Davies et al. found no evidence
for tidal streams which
could be associated with the sample of dEs.\\
A recent paper by \cite{mastropietro05} shows that the size of
harassed dEs could be smaller and streams surface brightness
fainter than what originally predicted. However, the most
problematic part of the harassment idea is where the initial LSB
disk galaxies come from. As galaxy groups come together to form
clusters this population should reside in the groups, while large
numbers of LSB disc galaxies are just not found in galaxy groups.

\item {\bf Tidal Dwarfs -} Slow speed close tidal interactions
between galaxies in clusters result in gas and stars being pulled
out from the interacting galaxies into giant streams, along which
clumps of gas and stars form. Over time the stream fades, and the
clump is classified as a Tidal Dwarf Galaxy (TDG). There is
observational evidence of the extistence of this kind of galaxies
(see e.g. \cite{hunsberger96}) and simulations of the formation of
TDGs predict that at most 1-2 of them form with each interaction
(\cite{okazaki00}, \cite{duc07}). We can estimate how many TDG
producing interactions there could be in Virgo by considering a
simple rate equation. The number of interactions (N) which may
produce a TDG in a cluster depends upon four parameters - the
number density of galaxies ($\rho$), their interaction cross
section ($\sigma$), their velocity (v) and the age of the cluster
(T). Thus,

\begin{equation}
N \sim \rho \sigma v T
\end{equation}

If we assume that only interactions between disk galaxies (S0 and
spirals) produce TDGs (\cite{okazaki00}), then using the
information in \cite{tully96} and assuming that the interaction
cross-section is the virial radius, we estimate there to be about
13 interactions per Gyr. So, if we assume that each interaction
makes, at most, 2 TDGs, this means that we would expect 26 TDGs to
be formed every Gyr. It seems therefore extremely unlikely that
TDGs make up a large fraction of the cluster dwarf galaxy
population, and certainly not large enough to account for the
dwarf galaxy population we find in Virgo today.

\item {\bf Ram pressure stripping -} Galaxies infalling in a
cluster with a hot intra-cluster medium are subject to its ram
pressure that can be capable of stripping thier gas away. dIrrs
could be particularly fragile with respect to this mechanism
(\cite{vanzee04}) and could potentially be transformed to dEs.
However, the importance of ram pressure stripping on the evolution
of cluster dwarf galaxies in Virgo was investigated by
\cite{sabatini05} who found that, due to the expected high M/L
ratio of these galaxies, only those dwarfs within the cluster core
($\sim$0.5Mpc or 1.5$^{\circ}$) would be affected by this process.
For the E-W strip, they also conclude that the dwarfs they detect
within the projected cluster core would be severely tidally
disrupted if they were actually located in the core, thus they
must be outside the core region, and therefore will not be subject
to ram pressure stripping. The majority (99\%) of the galaxies
detected in our N-S Virgo cluster strip are outside the projected
core region due to the offset of this strip from the cluster
centre. Thus the effect of ram pressure stripping
on these galaxies must be small. \\

In \cite{sabatini05} we suggest that enhanced star formation
triggered by interactions with the cluster and galaxy potentials,
accelerates the evolution of infalling DM halos so that they
resemble the dEs which we see in Virgo today, a process that does
not happen in the field. The (B-I) colour of our detections
(discussed in section \ref{sec:results}) are consistent with a
stellar population that is younger than the giant elliptical
galaxies.

\end{itemize}
\subsection{Suppression of dwarfs}
Having discussed the possible mechanisms which could create the
large population of dwarf galaxies that we find in our survey of
the Virgo cluster region, we now investigate the possible reasons
why we detect very few dwarf galaxies in lower density
environments (the general field, the region around M101 and the
low density Ursa Major cluster).
\begin{itemize}
\item {\bf Supernovae winds -} In this scenario (\cite{dekel86}),
the most commonly invoked when attempting to suppress the
formation of dwarf galaxies, the first generation of SN injects
enough energy into the halo gas, for it to escape the halo and
thereby prevent further star formation, rendering the halo
invisible. \cite{babul92} have suggested that this mechanism for
gas expulsion may be environmentally dependent because the
pressure of the intra-cluster gas will reduce gas loss in clusters
(pressure confinement). However, it is not clear how efficient
this mechanism for gas loss will be in any environment:
\cite{maclow99} have investigated this idea using numerical
simulations and showed that SN winds are effective at blowing out
the gas only for haloes of masses $<$10$^{6}$$M_{\odot}$.
\cite{sabatini05} have also questioned the viability of this gas
loss mechanism in the light of the very high mass-to-light ratios
that have been derived for some dwarf galaxies. Note also that,
using ASCA X-ray survey temperature maps, \cite{roberts07b} showed
that galaxies in Virgo are pressure confined only within 3 deg of
M87 in the NS strip and 2 deg in the EW. Thus gas expulsion via SN
winds does not appear to be able to explain why we see small
numbers of dwarf galaxies only in low density environments.

\item {\bf Re-ionization -} Are there many DM haloes present in
low density environments that are not observable because they have
not been lit up by star formation? One explanation for this is the
presence of a photoionizing background preventing the gas in the
halo from cooling. To account for an environmental dependence of
this mechanism, we should refer to the so called `squelching'
scenario (\cite{tully02}): in this picture high density cluster
sized regions (like the Virgo Cluster) are assumed to form before
the epoch of re-ionization (thus star formation in their dwarfs is
not inhibited). Lower density regions (such as Ursa Major and the
field) form later and thus the UV background heats the gas in the
dwarfs residing in them, preventing it from cooling and forming
stars. However, in their model, \cite{tully02} used $z_{reion}$ of
6, while the third year of WMAP results have pushed the epoch of
re-ionization to $z=9 \pm 3$, a time when the formation of dwarf
galaxy sized objects is rare.\\
Although the squelching scenario may have problems explaining the
environmental dependence of dwarf galaxy populations, the effect
of photoionization on low mass DM haloes may well play a part in
the formation of galaxies in the idea known as `downsizing'
(\cite{cowie96}). This scenario, born out of observational
evidence that larger galaxies have older stellar populations than
lower mass ones is at first sight contrary to hierarchical theory
of structure formation. However, it is not contrary if for some
reason star formation in low mass halos is in some way delayed,
possibly delayed so long that large numbers of small halos have
not yet undergone any star formation at all. \\
If photoionization does result in there being many low mass DM
haloes in the Universe which have not been able to form stars to
make them visible as dwarf galaxies, then gravitational lensing
could be used as a probe of substructure. This is an ideal tool to
use since light is deflected gravitationally by matter, whether it
is light or dark, thus if there were small dark haloes present in
the Universe, they could be detected by this means. Such studies
have been carried out (\cite{metcalf02}, 2002, \cite{bradac02},
\cite{dalal02}) and preliminary results show evidence for the
presence of substructure. \cite{kochanek03} also rule out the
possibilities of other effects causing the flux anomalies in a
further study of their data, concluding that `{\it{low mass haloes
remain the best explanation of the phenomenon'}}. However, if
these low mass DM haloes do exist in the numbers predicted by
$\Lambda$CDM, then as they fall through the disk of their parent
galaxy, they should heat the disk and cause it to thicken
(\cite{toth92}). This is contrary to some observations of old thin
disk systems or galaxies with no thick disk components, although
it is now being argued that the amount of heating and thickening
has been overestimated (\cite{font01}, \cite{velazquez99}). This
is clearly a matter for further investigation.

\end{itemize}
\section{Conclusions}
Our suggested solution to the origin of the differences in the
number counts and properties of dwarf galaxies in different
environments is that there must be many very LSB or totally dark
galaxies in the Universe that we have not yet been discovered. In
the cluster environment many of these have been 'lit up' by
enhanced star formation due to them being pulled and pushed around
within the cluster environment. In support of this we cite the
following results from our survey:
\begin{enumerate}
\item The dE galaxies in Virgo are bluer than the giant
ellipticals - their star formation was delayed until the cluster
was formed. \item Dwarf galaxies in Virgo are redder and gas poor
compared to those in the field - the cluster environment promotes
the conversion of gas into stars. Note that downsizing implies low
efficiency star formation in the lowest mass objects - but more
rapid in clusters? \item The galaxies we detect in Virgo are too
small to be the result of harassment. \item There are too few
tidal interactions in Virgo for them to be created tidally. \item
There is a clear lack of dwarf galaxies in the dynamically young
Ursa Major cluster. \item If the dwarfs have high mass-to-light
ratios they will not be subject to gas lose by SN driven winds.\\
\end{enumerate}

Multi-band follow ups to try and constrain the stellar population,
star formation history, age, metallicity of the dwarfs of our
catalogue are necessary in order to better investigate their
nature. We are currently using $ugriz$ SDSS imaging data at this
purpose for a subsample of them and we hope to obtain NIR data for
constraining their stellar population.\\

The recent discovery of extremely low luminosity and low surface
brightness dSph companions to the MW (\cite{irwin07} and
references therein) has highlighted the possibility that the
predicted population of low mass haloes in $\Lambda$CDM may
actually exist. \cite{kleyna05} comment that this new dSph, which
has a M/L of over 500M$_{\odot}/L_{\odot}$ and absolute magnitude,
$M_{V}$$\sim$-6.75, {\it{`may represent the best candidate for a
``missing'' $\Lambda$CDM halo'}}. They conclude that there must be
more dark and massive dwarfs hiding in the region around the Milky
Way. It is therefore extremely important that searches for such
objects are carried out if we are to properly check the
consistency of observations with $\Lambda$CDM predictions.


\begin{discussion}

\discuss{Cortese}{I am wondering if in comparing the contribution
of dwarf galaxies in different environments you have considered
that the field is underdense, i.e. the number of galaxies in the
field is small and therefore relative errors on counts are large.
Can the errors take into account of the discrepancy and reconcile
the faint end slope of the Virgo Cluster and that of the Field?
Same comment on the DGR.}

\discuss{Sabatini}{Number counts in the faint part of the LF in
the field indeed were so low that we couldn't do a LF. This is why
we have used the DGR for the comparison - the DGR allows to
rescale taking into account the different density of the
environments. However, indeed relative errors in counts are larger
in the field, but I don't think they could be responsible for
reconciling the discrepancy: we would have to have missed an order
of magnitude in number of dwarfs in the field compared to the
centre of the cluster.}

\discuss{Moore}{What about Valotto \etal\ 2001?}

\discuss{Sabatini}{Valotto \etal\ show that using statistical
subtraction for the estimation of the background contamination,
can lead to artificially steep LFs. For this reason, we do not use
control fields for a statistical estimation of the background. We
instead run numerical simulations to model the background galaxy
population and find the best selection criteria (based on
morphology) that allows to minimize the background contamination.
A far away galaxy looks faint and could in principle be a
contaminant in the cluster LF, but selecting galaxies for their
scale length and surface brightness ensures us to get rid of these
contaminants, that would appear smaller and with higher surface
brightness at a given magnitude if compared to local LSB galaxies}




\end{discussion}

\begin{thebibliography}{}
\bibitem[Babul \& Rees (1992)]{babul92}
    {Babul, A. \& Rees,M.,} 1992,
    \textit{MNRAS}, 255, 346
\bibitem[Bertin\& Arnouts (1996)]{bertin96}
    {Bertin, E., Arnouts, A.,} 1996,
    \textit{A $\&$ A}, 117, 393
\bibitem[Binggeli \etal\ (1984)]{binggeli84}
    {Binggeli, B., Sandage, A., Tarenghi, M.,} 1984,
    \textit{AJ}, 89, 64
\bibitem[Blanton \etal\ (2001)]{blanton01}
    {Blanton, M.R., Dalcanton, J., Eisenstein, D., Loveday, J., Strauss,
    M., SubbaRao, M., Weinberg, D.H., Anderson, J.E. Jr. and 61
    coauthors,}2001,
    \textit{AJ},121, 2358
\bibitem[Blanton \etal\ (2005)]{blanton05}
    {Blanton, M.R., Eisenstein, D., Hogg, D.W., Schlegel, D.J. \& Brinkmann,
    J.,}2005,
    \textit{ApJ}, 631, 208
\bibitem[Bradac \etal\ (2002)]{bradac02}
    {Bradac, M., Schneider, P., Steinmetz, M., Lombardi, M., King, L.
    \& Porcas, R.,} 2002,
    \textit{AA}, 388, 373
\bibitem[Conture \etal\ (1990)]{conture90}
    {Conture, J., Harris, W. \& Allwright, J.,} 1990,
    \textit{ApJS}, 73, 671
\bibitem[Cowie \etal\ (1996)]{cowie96}
    {Cowie, L., Songaila, A., Hu, E. \& Cohen, J.,} 1996,
    \textit{AJ}, 112, 839
\bibitem[Dalal \& Kochanek (2002)]{dalal02}
    {Dalal, N. \& Kochanek C.,} 2002,
    \textit{ApJ}, 572, 25
\bibitem[Davies \etal\ 2005]{davies05}
    {Davies, J.I., Roberts, S. \& Sabatini, S.,} 2005,
    \textit{MNRAS}, 356, 794
\bibitem[Dekel \& Silk (1986)]{dekel86}
    {Dekel, A. \& Silk, J.,} 1986,
    textit{ApJ}, 303, 39
\bibitem[Duc \etal\ (2007)]{duc07}
    {Duc, P.A., Bournaud, F. \& Boquien, M.,} 2007
    \textit{IAUS}, 237, 323
\bibitem[Font \etal\ (2001)]{font01}
    {Font, A., Navarro, J., Stadel, J. \& Quinn, T.,} 2001,
    \textit{ApJ}, 563, L1
\bibitem[Hunsberger \etal\ (1996)]{hunsberger96}
    {Hunsberger, S., Charlton, J. \& Zaritsky, D.,} 1996,
    \textit{ApJ}, 462, 50
\bibitem[Irwin \etal\ (2007)]{irwin07}
    {Irwin, M.J., Belokurov, V., Evans, N.W., Ryan-Weber, E.V., de Jong, J.T.A.,
    Koposov, S., Zucker, D.B., Hodgkin, S.T. and 19
    coauthors,},2007,
    \textit{ApJ}, 656L, 13
\bibitem[Jerjen \etal\ (2004)]{jerjen04}
    {Jerjen, H., Binggeli, B., Barazza, F.,} 2004,
    \textit{AJ}, 127, 771
\bibitem[Kambas \etal\ (2000)]{kambas00}
    {Kambas, A., Davies, J.I., Smith, R., Bianchi, S., Haynes
    J.A.,}2000,
    \textit{AJ}, 120, 1316
\bibitem[Kauffman \etal\ (1993)]{kauffman93}
    {Kauffman, G., White S.D.M. \& Guiderdoni B.,} 1993,
    \textit{MNRAS}, 264, 201
\bibitem[Kleyna \etal\ (2005)]{kleyna05}
    {Kleyna, J., Wilkinson, M.I., Evans, N.W., Gilmore, G., Frayn, C.,} 2005,
    \textit{ApJ}, 630, 141
\bibitem[Klypin \etal\. (1999)]{klypin99}
    {Klypin, A., Kravtsov, A.V., Valenzuela O. \& Prada, F.,}
    1999, \textit{ApJ}, 522, 82
\bibitem[Kochanek \& Dalal (2003)]{kochanek03}
    {Kochanek, C. \& Dalal, N.,} 2003,
    \textit{AIP Conf. Proc. 666}, 'The Emergence of Cosmic Structure', p. 103
\bibitem[Liske \etal\ (2003)]{liske03}
    {Liske, J., Lemon, D., Driver, S., Cross, N. \& Couch, W.,} 2003,
    \textit{MNRAS}, 344, 307
\bibitem[MacLow \& Ferrara (1999)]{maclow99}
    {MacLow, M. \& Ferrara, A.,} 1999,
    \textit{ApJ}, 513, 142
\bibitem[Mastropietro \etal\ (2005)]{mastropietro05}
    {Mastropietro, C., Moore, B., Mayer, L., Debattista, V.P., Piffaretti, R., Stadel,
    J.,}, 2005,
    \textit{MNRAS},364, 607
\bibitem[Metcalf \& Zhao (2002)]{metcalf02}
    {Metcalf, R. \& Zhao, H.,} 2002,
    \textit{ApJ}, 567, L5
\bibitem[Michard (2000)]{michard00}
    {Michard, R.,} 2000,
    \textit{AA}, 360, 85
\bibitem[Moore \etal\ (1996)]{moore96}
    {Moore, B., Katz, N., Lake, G., Dressler, A. \& Olmer,
    A.,}1996,
    \textit{Nature}, 379, 613
\bibitem[Moore \etal\ (1999)]{moore99}
    {Moore, B., Ghigna, S., Governato, F., Lake, G., Quinn, T.,
    Stadel, J. \& Tozzi, P.,} 1999, \textit{ApJ}, 524, L19
\bibitem[Norberg \etal\ (2002)]{norberg02}
    {Norberg, P., Cole, S., Baugh, C.M., Frenk, C.S., Baldry, I., Bland-Hawthorn, J., Bridges, T., Cannon, R.,
     and 20 coauthors,}2002,
     textit{MNRAS}, 336, 907
\bibitem[Okazaki \& Taniguchi (2000)]{okazaki00}
    {Okazaki, T. \& Taniguchi, Y.,} 2000,
    \textit{ApJ}, 543, 149
\bibitem[Roberts \etal\ (2004)]{roberts04}
    {Roberts, S., Davies, J.I., Sabatini, S., van Driel, W., O'Neil, K.,
      Baes, M., Linder, S.M., Smith, R., Evans, R.,} 2004,
      \textit{MNRAS}, 352, 478
\bibitem[Roberts (2005)]{roberts07b}
    {Roberts, S.,} 2005,
    \textit{PhD Thesis}, Cardiff University
\bibitem[Roberts \etal\ (2007)]{roberts07}
    {Roberts, S., Davies, J.I., Sabatini, S., Auld, R., Smith,
    R.,}2007
    \textit{MNRAS}, in press
\bibitem[Sabatini \etal\ (2003)]{sabatini03}
    {Sabatini, S., Davies, J.I., Scaramella, R., Smith, R., Baes, M.,
    Linder, S.M., Roberts S. \& Testa, V.,} 2003, \textit{MNRAS},
    341, 981
\bibitem[Sabatini \etal\ (2005)]{sabatini05}
    {Sabatini, S., Davies, J.I., Van Driel, W., Baes, M., Robert, S.,
    Smith, R., Linder, S.,O'Neil, K.,} 2005,
    \textit{MNRAS}, 357, 819
\bibitem[Simon \& Geha (2007)]{simon_geha07}
    {Simon, J.D. \& Geha M.,} 2007, submitted to
    \textit{ApJ} (preprint: astro-ph/0706.0516v1)
\bibitem[T{\'o}ht \& Ostriker (1992)]{toth92}
    {Toth, G. \& Ostriker, J.,} 1992,
    \textit{ApJ}, 389, 5
\bibitem[Tully \etal\ (1996)]{tully96}
    {Tully, B.R., Verheijen, M.A.W., Pierce, M.J., Huang, J.S., Wainscoat,
    R.J.,}1996,
    \textit{AJ}, 112, 2471
\bibitem[Tully \etal\ (2002)]{tully02}
    {Tully, B.R., Somerville, R., Trentham, N. \& Verheijen, M.,} 2002,
    \textit{ApJ}, 569, 573
\bibitem[van Zee \etal\ (2004)]{vanzee04}
    {van Zee, L., Barton, E.J. \& Skillman, E.D.,} 2004,
    \textit{AJ}, 128, 2797
\bibitem[Vel{\'a}zquez \& White (1999)]{velazquez99}
    {Vel{\'a}zquez, H. \& White, S.,} 1999,
    \textit{MNRAS}, 304, 254
\bibitem[Zwaan \etal\ (2005)]{zwaan05}
    {Zwaan, M.A., Stavely-Smith, L., Koribalski, B.S., Henning, P.A.,
    Kilborn, V.A., Ryder,, S.D., Barnes, D.G., Bhathal Rl \etal\,}
    2005, \textit{AJ}, 1255, 2842

\end{thebibliography}
\end{document}